\documentclass[preprintnumbers,
amsmath,amssymb,floatfix,10pt,prd,onecolumn,
superscriptaddress,longbibliography,nofootinbib]{revtex4-2}
\usepackage{booktabs}
\usepackage{multirow}
\usepackage{bm}
\usepackage{amsfonts}
\usepackage{latexsym}
\usepackage{amsmath}
\usepackage{textcomp}
\usepackage{float}
\usepackage{booktabs}
\usepackage{dcolumn}
\usepackage{dsfont}
\usepackage{ragged2e}
\usepackage{epsfig}
\usepackage[dvipsnames]{xcolor}
\usepackage{hyperref}
\hypersetup{           
	colorlinks=true,                
	breaklinks=true,                
	urlcolor= OrangeRed,                
	linkcolor= magenta,                
	bookmarksopen=false,
	filecolor=black,
	citecolor=red,
	linkbordercolor=blue}
\usepackage{graphicx}
\usepackage{orcidlink}
\usepackage[T1]{fontenc}

\begin{document}
	\title{Constraint on Symmetric Teleparallel Gravity with Different Dark energy Parametrizations from DESI DR2 BAO Data}
	
	\author{Rajdeep Mazumdar \orcidlink{0009-0003-7732-875X}}%
	\email{rajdeepmazumdar377@gmail.com}
	\affiliation{%
		Department of Physics, Dibrugarh University, Dibrugarh \\
		Assam, India, 786004}

	\author{Mrinnoy M. Gohain\orcidlink{0000-0002-1097-2124}}
	\email{mrinmoygohain19@gmail.com}
	\affiliation{%
		Department of Physics, Dibrugarh University, Dibrugarh \\
		Assam, India, 786004}
	
	\author{Kalyan Bhuyan\orcidlink{0000-0002-8896-7691}}%
	\email{kalyanbhuyan@dibru.ac.in}
	\affiliation{%
		Department of Physics, Dibrugarh University, Dibrugarh \\
		Assam, India, 786004}%
	\affiliation{Theoretical Physics Divison, Centre for Atmospheric Studies, Dibrugarh University, Dibrugarh, Assam, India 786004}
	
\begin{abstract}
We investigate the cosmological viability of symmetric teleparallel gravity, specifically the $f(Q)$ gravity model with a power-law form $f(Q) = \alpha Q^n$, in combination with two widely used dark energy parameterizations: Chevallier–Polarski–Linder (CPL) and Barboza–Alcaniz (BA). Employing the most recent DESI DR2 Baryon Acoustic Oscillation (BAO) dataset along with previous BAO measurements, we constrain the model parameters through a robust Markov Chain Monte Carlo (MCMC) analysis. We examine the background evolution via key cosmological indicators including the Hubble parameter $H(z)$, deceleration parameter $q(z)$, the effective equation of state $\omega_{\rm eff}(z)$, and the Om diagnostic. Our results indicate that the inclusion of DESI DR2 data significantly tightens constraints on the model parameters and supports a consistent transition from decelerated to accelerated expansion. The present-day values of $q(0)$ and $\omega_{\rm eff}(0)$ lie within the quintessence regime for all datasets. However, for lower redshift values the behaviour varies between phantom-like and quintessence-like phases. Statistical comparison via $\chi^2$, AIC, BIC, and $R^2$ further demonstrate that both CPL + $f(Q)$ and BA + $f(Q)$ provide better or competitive fits to data compared to $\Lambda$CDM, hence offering a compelling geometric alternative for explaining late-time cosmic acceleration.
\end{abstract}

	
	\maketitle
\section{Introduction}\label{sec1}
Recent advancements in observational cosmology—including Type Ia supernovae (SN Ia) surveys~\cite{riess1998,perlmutter1999}, Cosmic Microwave Background (CMB) measurements~\cite{bennett2003,spergel2003}, Baryon Acoustic Oscillations (BAO)~\cite{eisenstein2005,percival2010}, and Large Scale Structure (LSS) data—have provided compelling and consistent evidence that the Universe is currently experiencing an accelerated expansion phase. This late-time acceleration is widely attributed to an enigmatic component known as dark energy (DE), which is estimated to comprise approximately 70\% of the total energy density of the Universe. Despite its success in explaining observations, the fundamental nature of DE remains theoretically ambiguous. The prevailing cosmological paradigm, the $\Lambda$ Cold Dark Matter ($\Lambda$CDM) model, interprets this acceleration via a cosmological constant $\Lambda$. However, this interpretation is not without difficulties, notably the so-called fine-tuning problem and the coincidence problem~\cite{weinberg1989,carroll1992,sahni2000}, which continue to motivate the exploration of alternative explanations. This ongoing quest to explain late-time cosmic acceleration has led to two broad theoretical strategies: one that introduces new exotic energy components—such as quintessence, phantom fields, k-essence~\cite{ArmendarizPicon2000}, tachyon~\cite{Sen2002}, chameleon~\cite{Khoury2004}, and Chaplygin gas models~\cite{Kamenshchik2001,Bento2002}—and another that modifies the underlying theory of gravity itself. Among the various modified gravity frameworks, a relatively recent and actively explored approach is known as symmetric teleparallel gravity, or \( f(Q) \) gravity~\cite{Jimenez2018, BeltranJimenez2019}.\\
Unlike General Relativity, which is built on curvature, and teleparallel gravity, which is based on torsion, the \( f(Q) \) theory constructs gravitational dynamics using the non-metricity scalar \( Q \). In the so-called coincident gauge, the affine connection is chosen such that the torsion and curvature vanish, leaving the non-metricity as the sole geometric quantity encoding gravitational effects. This alternative geometrical formulation allows for the emergence of cosmic acceleration without invoking a cosmological constant or scalar fields, making it a compelling candidate for addressing both theoretical and observational challenges in modern cosmology. Recent studies have shown that \( f(Q) \) gravity models can successfully reproduce cosmological observations, including Supernovae Type Ia (SNIa)~\cite{Anagnostopoulos2021}, Cosmic Microwave Background (CMB) data~\cite{Lazkoz2021}, and Baryon Acoustic Oscillation (BAO) measurements~\cite{Capozziello2022}. These findings suggest that \( f(Q) \) gravity can closely mimic, or even outperform, the predictions of the standard $\Lambda$CDM model under certain conditions. Consequently, there is growing interest in examining its cosmological viability through various functional forms of \( f(Q) \). Several specific investigations underscore the versatility of this framework. Solanki et al.~\cite{Solanki2022} explored bulk viscous cosmology under different forms of \( f(Q) \); Hassan et al.~\cite{Hassan2022} examined the influence of the Generalized Uncertainty Principle (GUP) on Casimir wormholes within \( f(Q) \) gravity. Large-scale structure formation has also been investigated in this context~\cite{LSS2023}. Koussour et al.~\cite{Koussour2023} performed a late-time analysis using a power-law form \( f(Q) = \alpha Q^n \) with a constant sound speed, while Gadbail et al.~\cite{Gadbail2023} demonstrated the possibility of a transition from decelerated to accelerated expansion via a hybrid Chaplygin gas model with bulk viscosity in power-law \( f(Q) \) gravity. Other proposals, such as parametrized Hubble expansions~\cite{KoussourParam2023}, further enrich the landscape of viable models within this theory. Taken together, these results establish \( f(Q) \) gravity as a promising alternative to General Relativity and $\Lambda$CDM, motivating further study using evolving dark energy parametrizations and the latest cosmological datasets.\\
In the literature, the equation of state (EoS) parameter is widely employed as a diagnostic tool to characterize the nature of dark energy (DE) in various cosmological models. The EoS parameter, defined as the ratio of pressure \( p \) to energy density \( \rho \), encapsulates the dynamical behaviour of different energy components in the Universe. Its value varies depending on the dominant constituent: for instance, \( \omega = 0 \) corresponds to pressure-less matter, \( \omega = 1/3 \) characterizes radiation, and \( \omega = -1 \) represents a cosmological constant. For an accelerating Universe, the EoS parameter typically satisfies \( \omega < -1/3 \). The range \( -1 < \omega < -1/3 \) describes quintessence-like DE, while \( \omega < -1 \) is indicative of phantom energy~\cite{Caldwell2002, Carroll2003}. The most recent constraints from the Planck Collaboration suggest a current value of \( \omega_0 = -1.028 \pm 0.032 \)~\cite{Planck2018, PlanckDE2018}, hinting at a mild preference for a phantom regime. To model the evolution of dark energy in a flexible, phenomenological, and model-independent manner, several parametrizations of the EoS have been proposed in the literature. Among them the most widely used are:
\begin{itemize}
    \item \textbf{Chevallier–Polarski–Linder (CPL)} parametrization~\cite{Chevallier2001, Linder2003}:
    \begin{equation}
        \omega(z) = \omega_0 + \omega_1 \frac{z}{1 + z}
        \label{ps1}
    \end{equation} 
    \item \textbf{Barboza–Alcaniz (BA)} parametrization~\cite{Barboza2008}:
    \begin{equation}
        \omega(z) = \omega_0 + \omega_1 \frac{z(1 + z)}{1 + z^2}
        \label{ps3}
    \end{equation}
\end{itemize}
Both parametrizations provide redshift-dependent forms of the dark energy equation of state \(\omega(z)\) that reduce to the present-day value \(\omega_0\) at \(z=0\). The CPL parametrization evolves linearly with scale factor and approaches a constant value at high redshift, making it a simple yet effective tool for cosmological fits. The BA parametrization, with its rational function form, offers more flexibility to model potential deviations from \(\Lambda\)CDM, especially at intermediate redshifts. Conclusively, these parametrizations are particularly useful to model dark energy in a general and evolving framework, followed by parametric freedom for efficient comparisons with observational data.\\ 
Recent data releases from the DESI survey, particularly the DESI DR2 dataset, have emerged as powerful tools for imposing stringent constraints on a wide array of cosmological models \cite{refk57}. Owing to the exceptional precision of its Baryon Acoustic Oscillation (BAO) measurements, DESI has enabled detailed testing of numerous extensions beyond the standard $\Lambda$CDM framework. Notably, this dataset has been instrumental in constraining dynamical dark energy models \cite{refk58,refk59,refk60,refk61,refk62,refk63,refk64}, early dark energy scenarios \cite{refk65}, and a range of scalar field theories with both minimal and non-minimal couplings \cite{refk66,refk67,refk68}. In addition, DESI BAO data have been applied to explore quantum-gravity-inspired models, such as those based on the Generalized Uncertainty Principle \cite{refk69}, as well as interacting dark sector frameworks \cite{refk70,refk71,refk72}. Other applications include astrophysical tests \cite{refk73}, model-independent cosmographic reconstructions \cite{refk74}, and investigations into a wide spectrum of modified gravity and entropy-based theories \cite{refk62,refk75,refk76,refk77,refk78}, alongside various alternative cosmological scenarios (consult Ref.\cite{refk79} to Ref.\cite{refk100}). \\
In this work, we incorporate two different forms dark energy parametrizations schemes (CPL and BA) within the framework of \( f(Q) \) gravity. Particularly considering a power-law form of the non-metricity function, namely \( f(Q) = \alpha Q^n \), to examine their constraints and cosmological dynamics by employing Baryon Acoustic Oscillation (BAO) data from both pre-DESI or previous BAO (from observations such as SDSS and WiggleZ) and recent DESI DR2 BAO releases. In each case Markov Chain Monte Carlo (MCMC) is performed to analyse the constrain on the parameters, and investigation of the resulting cosmological behaviour through key indicators such as the deceleration parameter \( q(z) \), and effective equation of state \( w_{\text{eff}}(z) \) is conducted. For a review of previous works under the same framework one can consult Ref.\cite{x1} to Ref.\cite{x5} and other references within them. Our goal is to assess the viability of symmetric teleparallel gravity with evolving DE, and to investigate whether the inclusion of DESI data leads to tighter constraints or new cosmological features.\\
The structure of this paper is as follows: Sec. \ref{sec2} provides a brief mathematical formalism of $f(Q)$ gravity in a flat FLRW Universe, Sec. \ref{sec3} provides cosmological framework regarding the specific $f(Q)$ gravity model, and the derived cosmological parameters using the parametrization schemes. Sec. \ref{sec4} discusses the observational data and methodology used in the study. Sec \ref{sec5} discuses the constraints on the models and the cosmological implications it presents. Finally, Sec. \ref{sec6} provides a summary of the conclusions drawn from the study.
\section{$f(Q)$ Gravity Theory}\label{sec2}
In the language of differential geometry, the spacetime metric $g_{\mu\nu}$ defines distances, angles, and volumes, whereas the affine connection $\Upsilon^\gamma_{\mu\nu}$ governs parallel transport and covariant derivatives. In $f(Q)$ gravity, the connection is assumed to be curvature- and torsion-free, but with non-vanishing non-metricity:
\begin{equation}
Q^\gamma_{\mu\nu} \equiv -\nabla_\gamma g_{\mu\nu}.
\end{equation}
This non-metricity quantifies how the length of a vector changes under parallel transport. The affine connection can be decomposed as:
\begin{equation}
\Upsilon^\gamma_{\mu\nu} = \Gamma^\gamma_{\mu\nu} + L^\gamma_{\mu\nu},
\end{equation}
where $\Gamma^\gamma_{\mu\nu}$ is the Levi-Civita connection, and $L^\gamma_{\mu\nu}$ is the disformation tensor constructed from the non-metricity tensor. The non-metricity scalar $Q$ is constructed from contractions of the non-metricity tensor and a superpotential tensor $P^\gamma_{\mu\nu}$, defined as:
\begin{equation}
Q = -Q^\gamma_{\mu\nu} P^\mu_\gamma{}^\nu.
\end{equation}
The action for $f(Q)$ gravity in the presence of matter fields is given by:
\begin{equation}
S = \int d^4x \sqrt{-g} \left[ -\frac{1}{2} f(Q) + \mathcal{L}_m \right],
\end{equation}
where $f(Q)$ is a general function of the non-metricity scalar and $\mathcal{L}_m$ denotes the matter Lagrangian. Varying the action with respect to the metric yields the modified gravitational field equations:
\begin{equation}
2 \nabla_\gamma \left( \sqrt{-g} f_Q P^\gamma_{\mu\nu} \right) + \sqrt{-g} \left[ \frac{1}{2} f g_{\mu\nu} + f_Q \left(P_{\nu\rho\sigma} Q_\mu{}^{\rho\sigma} - 2 P_{\rho\sigma\mu} Q^\rho{}_{\nu}{}^\sigma \right) \right] = \sqrt{-g} T_{\mu\nu},
\end{equation}
where $f_Q = df/dQ$ and $T_{\mu\nu}$ is the energy-momentum tensor. In the coincident gauge, where the connection vanishes ($\Upsilon^\gamma_{\mu\nu} = 0$), the spacetime becomes flat and simplifies calculations. In this gauge, the non-metricity tensor becomes $Q^\gamma_{\mu\nu} = -\partial_\gamma g_{\mu\nu}$, making the covariant derivatives reduce to partial derivatives.\\
To explore cosmological implications, we assumes a spatially flat Friedmann–Lemaître–Robertson–Walker (FLRW) metric:
\begin{equation}
ds^2 = -dt^2 + a^2(t)\left( dx^2 + dy^2 + dz^2 \right),
\end{equation}
where $a(t)$ is the scale factor. For this metric, the non-metricity scalar becomes:
\begin{equation}
Q = 6H^2,
\end{equation}
with $H = \dot{a}/a$ being the Hubble parameter. We consider the energy-momentum tensor, given as $T_{\mu\nu} = (\rho + p)u_\mu u_\nu + p g_{\mu\nu}$, where $u^\mu = (-1,0,0,0)$ is the fluid's four-velocity that meets the condition $u^\mu u_\mu = -1$, $H$ stands for the Hubble parameter, and $\rho$ and $p$ represent the energy density and pressure, respectively. Using a perfect fluid energy-momentum tensor and the above metric, the modified Friedmann equations in $f(Q)$ gravity become:
\begin{equation}
3H^2 = \frac{1}{2f_Q} \left( \rho + \frac{f}{2} \right), \qquad \dot{H} + 3H^2 + \frac{\dot{f}_Q}{f_Q} H = \frac{1}{2f_Q} \left( -p + \frac{f}{2} \right).
\end{equation}
We have assumed $8\pi G =1$, where the derivative with respect to time is indicated by the overhead dot (.).  These equations describe how the expansion of the universe is impacted by the cosmological constant, space curvature, and the energy density of matter and radiation. If we set $f(Q) = Q$, the standard GR equations are recovered. To isolate the effective dark energy sector purely due to the geometry, we move the matter contributions to one side and define:
\begin{equation}
3H^2 = \rho_m + \rho_r + \rho_{de}, \qquad \dot{H} + 3H^2 = -\frac{1}{2}(\rho_m + \rho_r + \rho_{de} + p_{de}).
\end{equation}
Here, $\rho_r$, $\rho_m$, $p_m$, and $p_r$ represent the energy densities of the radiation and matter components, with $p_m$ and $p_r$ indicating the pressure associated with matter and radiation, respectively. Along with the effective dark energy density $\rho_{de}$ and pressure $p_{de}$ arising from the geometric contribution can be written as:
\begin{equation}
\rho_{de}(Q) = \frac{1}{2f_Q} \left( \frac{f(Q)}{2} - Q f_Q \right), \qquad
p_{de}(Q) = -\frac{1}{2f_Q} \left( \frac{f(Q)}{2} - Q f_Q + 4 \dot{H} f_{QQ} Q \right),
\end{equation}
where $f_Q = \frac{df}{dQ}$ and $f_{QQ} = \frac{d^2f}{dQ^2}$. Since $Q = 6H^2$, we have $\dot{Q} = 12 H \dot{H}$, and hence the time derivative of $f_Q$ becomes:
\begin{equation}
\dot{f}_Q = f_{QQ} \dot{Q} = 12 H \dot{H} f_{QQ}.
\end{equation}
Therefore, the effective equation of state (EoS) parameter for the dark energy component in $f(Q)$ gravity is given by:
\begin{equation}
\omega_{de} = \frac{p_{de}}{\rho_{de}} = -1 + \frac{4 \dot{H} f_{QQ} Q}{\frac{f(Q)}{2} - Q f_Q}.
\end{equation}
Assuming non-interacting radiation, matter, and geometric dark energy components, their energy conservation equations are:
\begin{equation}
\dot{\rho}_r + 4H\rho_r = 0, \qquad \dot{\rho}_m + 3H\rho_m = 0, \qquad \dot{\rho}_{de} + 3H(1 + \omega_{de})\rho_{de} = 0.
\end{equation}
These can help to derive standard redshift evolution:
\begin{equation}
\rho_r(z) = \rho_{r0}(1 + z)^4, \qquad \rho_m(z) = \rho_{m0}(1 + z)^3.
\end{equation}
where $\rho_{r0}$ and $\rho_{m0}$ denote the presentvalues of $\rho_{r}$ and $\rho_{r}$. Thus, $f(Q)$ gravity provides a powerful and consistent geometric framework for describing cosmic evolution, with effective dark energy behavior emerging from the non-metricity-based modification of spacetime geometry.
\section{Cosmological Framework} \label{sec3}
In this work, we consider a power-law form of the $f(Q)$ gravity model, given by:
\begin{equation}
f(Q) = \alpha Q^n,
\end{equation}
where $\alpha$ and $n$ are free parameters of the theory. This model generalizes the symmetric teleparallel equivalent of general relativity and has been explored in various cosmological scenarios, including linear ($n = 1$) and quadratic ($n = 2$) cases. For this choice of $f(Q)$, the effective dark energy density and pressure derived from the modified Friedmann equations are expressed as:
\begin{equation}
\rho_{de}(z) = \alpha \cdot 6^n \left( \frac{1}{2} - n \right) H^{2n}(z),
\end{equation}
\begin{equation}
p_{de}(z) = -\alpha \cdot 6^{n-1} \left( \frac{1}{2} - n \right) H^{2(n-1)}(z) \left( 3H^2(z) + 2n\dot{H} \right).
\end{equation}
From these, the dark energy equation of state (EoS) parameter becomes:
\begin{equation}
\omega_{de}(z) = -1 - \frac{2n}{3} \left( \frac{\dot{H}}{H^2} \right).
\label{m1}
\end{equation}
To evaluate the time evolution of the Hubble parameter in terms of redshift, we use the relation:
\begin{equation}
\dot{H} = \frac{dH}{dt} = - (1 + z) H(z) \frac{dH}{dz}.
\label{m2}
\end{equation}
Using it we can obtain $H(z)$ which describes the rate of cosmic expansion as a function of redshift. Further we can obtain the deceleration parameter, given as:
\begin{equation}
    q(z) = -1 - \frac{\dot{H}}{H^2} = -1 + (1+z) \frac{1}{H(z)} \frac{dH}{dz},
\end{equation}
which indicates the acceleration or deceleration of the Universe. A negative value of $q(z)$ corresponds to accelerated expansion. Also, we can obtain the effective EoS, given as:
\begin{equation}
\omega_{eff}(z) = -1 + \frac{2(1+z)}{3H(z)}\frac{dH}{dz},
\end{equation}
which provides an effective fluid description of the Universe's dynamics, incorporating both matter and dark energy effects. These quantities allow us to examine whether the model predicts a transition from deceleration to acceleration and to test whether the effective equation of state crosses the phantom divide ($w_{\text{eff}} < -1$).\\
Noting that equations (for $\rho_{de}$ and $p_{de}$) form an underdetermined system (two equations with three unknowns: $H$, $\rho_{de}$, and $p_{de}$), we close the system by prescribing a parametrization for the dark energy EoS. We consider two widely studied dark energy parametrizations schemes: the Chevallier–Polarski–Linder (CPL) and Barboza–Alcaniz (BA) parametrizations as defined by
Eq. (\ref{ps1}) and  (\ref{ps3}). Using them with Eq. (\ref{m1}) and (\ref{m2}) we obtained the two cases:
\begin{itemize}
\item \textbf{CPL + $f(Q)$}: For which we have $H(z)$, $q(z)$, and $\omega_{eff}(z)$ as;\\
\begin{equation}
H(z)= H_0 e^{-\frac{3 \omega _1 z}{2 n z+2 n}} (z+1)^{\frac{3 \left(\omega _0+\omega _1+1\right)}{2 n}}
\label{hz1}
\end{equation}
\begin{equation}
q(z)= \frac{3 \left(\omega _0+\left(\omega _0+\omega _1+1\right) z+1\right)}{2 n (z+1)}-1
\label{qz1}
\end{equation}
\begin{equation}
\omega_{eff}(z)= \frac{-n (z+1)+\omega _0+\left(\omega _0+\omega _1+1\right) z+1}{n (z+1)}
\label{wz1}
\end{equation}
\item \textbf{BA + $f(Q)$}: For which we have $H(z)$, $q(z)$, and $\omega_{eff}(z)$ as;\\
\begin{equation}
H(z)= H_0 \left((z+1)^{\omega _0+1} \left(z^2+1\right)^{\frac{\omega _1}{2}}\right){}^{\frac{3}{2 n}}
\label{hz3}
\end{equation}
\begin{equation}
q(z)= \frac{3 \left(\omega _0+\left(\omega _0+\omega _1+1\right) z^2+\omega _1 z+1\right)}{2 n \left(z^2+1\right)}-1
\label{qz3}
\end{equation}
\begin{equation}
\omega_{eff}(z)= \frac{-n \left(z^2+1\right)+\omega _0+\left(\omega _0+\omega _1+1\right) z^2+\omega _1 z+1}{n \left(z^2+1\right)}
\label{wz3}
\end{equation}
\end{itemize}
For the above cases we try to constraint the parameters using DESI DR2 BAO data along with previous BAO (P-BAO) data in the following sections, and investigate the cosmological dynamics they implicate.

\section{Observational Data and Methodology}\label{sec4}
In order to constrain the parameters of the $f(Q)$ gravity model with various dark energy (DE) parametrizations, we employ a combination of Baryon Acoustic Oscillation (BAO) data from both pre-DESI or previous BAO (from observations such as SDSS and WiggleZ) and recent DESI DR2 BAO releases. Other cosmological probes such as Type Ia supernovae  or cosmic chronometers, are avoided here so as to solely focus on the constrains and cosmological implication presented by the introduction of the DESI DR2 BAO dataset in comparison to the previous BAO datasets. We analyze the data through a Markov Chain Monte Carlo (MCMC) framework to obtain robust constraints on the cosmological and model parameters. Under the light of which then the cosmological scenarios are studied using key cosmological indicators such as the deceleration parameter \( q(z) \), effective equation of state \( w_{\text{eff}}(z) \), statefinder diagnostics and $om$ diagnostics.\\
\begin{table}[h!]
\centering
\begin{tabular}{|ccc|c||ccc|c|}
\hline
\multicolumn{4}{|c||}{\textbf{DESI}} & \multicolumn{4}{c|}{\textbf{P-BAO}} \\
\hline
$z$ & $H(z)$ & $\sigma_H$ & Ref & $z$ & $H(z)$ & $\sigma_H$ & Ref \\
\hline
0.51 & 97.21 & 2.83 & \cite{ref98} & 0.24 & 79.69 & 2.99 & \cite{d113} \\
0.71 & 101.57 & 3.04 & \cite{ref98} & 0.30 & 81.70 & 6.22 & \cite{d114} \\
0.93 & 114.07 & 2.24 & \cite{ref98} & 0.31 & 78.17 & 6.74 & \cite{d115} \\
1.32 & 147.58 & 4.49 & \cite{ref98} & 0.34 & 83.17 & 6.74 & \cite{d113} \\
2.33 & 239.38 & 4.80 & \cite{ref98} & 0.35 & 82.70 & 8.40 & \cite{d116} \\
 &  &  &  & 0.36 & 79.93 & 3.39 & \cite{d115} \\
 &  &  &  & 0.38 & 81.50 & 1.90 & \cite{d5} \\
 &  &  &  & 0.40 & 82.04 & 2.03 & \cite{d115} \\
 &  &  &  & 0.43 & 86.45 & 3.68 & \cite{d113} \\
 &  &  &  & 0.44 & 82.60 & 7.80 & \cite{d74} \\
 &  &  &  & 0.44 & 84.81 & 1.83 & \cite{d115} \\
 &  &  &  & 0.48 & 87.79 & 2.03 & \cite{d115} \\
 &  &  &  & 0.56 & 93.33 & 2.32 & \cite{d115} \\
 &  &  &  & 0.57 & 87.60 & 7.80 & \cite{d10} \\
 &  &  &  & 0.57 & 96.80 & 3.40 & \cite{d117} \\
 &  &  &  & 0.59 & 98.48 & 3.19 & \cite{d115} \\
 &  &  &  & 0.60 & 87.90 & 6.10 & \cite{d74} \\
 &  &  &  & 0.61 & 97.30 & 2.10 & \cite{d5} \\
 &  &  &  & 0.64 & 98.82 & 2.99 & \cite{d115} \\
 &  &  &  & 0.978 & 113.72 & 14.63 & \cite{d118} \\
 &  &  &  & 1.23 & 131.44 & 12.42 & \cite{d118} \\
 &  &  &  & 1.48 & 153.81 & 6.39 & \cite{d79} \\
 &  &  &  & 1.526 & 148.11 & 12.71 & \cite{d118} \\
 &  &  &  & 1.944 & 172.63 & 14.79 & \cite{d118} \\
 &  &  &  & 2.30 & 224.00 & 8.00 & \cite{d119} \\
 &  &  &  & 2.36 & 226.00 & 8.00 & \cite{d120} \\
 &  &  &  & 2.40 & 227.80 & 5.61 & \cite{d121} \\
\hline
\end{tabular}
\caption{Observed Hubble parameter $H(z)$ (in units of $km s^{-1} Mpc^{-1}$) and their uncertainties at redshift $z$ form the DESI and P-BAO datasets.}
\label{table:4}
\end{table}\\
To find the mean values of the parameters $H_0$, $n$, $\omega_0$ and $\omega_1$, we employ the chi-squared minimization method using the following chi-squared function:
\begin{equation}
\chi^2_H(H_0,n,\omega_0,\omega_0) = \sum_{i=1}^{N} \frac{\left[ H_{th}(z_i, H_0,n,\omega_0,\omega_0) - H_{obs}(z_i) \right]^2}{\sigma_H(z_i)^2},
\label{32}
\end{equation}
where, $\sigma_H(z_i)$ denotes the standard error in the observed value of $H(z_i)$ at redshift $z_i$, $H_{th}$ represents the theoretical value of the Hubble parameter, and $H_{obs}$ represents the observed value. We conduct the MCMC analysis in three stages: using the DESI dataset alone, then with the previous BAO dataset, and finally with a combination of both. The outcomes are compared with those from the standard \(\Lambda\)CDM model using statistical indicators such as the coefficient of determination \( R^2 \), the minimum chi-squared value \( \chi^2_{\min} \), the Akaike Information Criterion (AIC), and the Bayesian Information Criterion (BIC) (see Refs.~\cite{refS66}, \cite{refS68}, \cite{refS64} for details). A model with lower \( \chi^2_{\min} \), AIC, and BIC values is statistically preferred, indicating a better trade-off between goodness of fit and complexity. Among these, \( R^2 \) and \( \chi^2_{\min} \) assess fit quality without accounting for the number of free parameters. In contrast, AIC and BIC penalize models with greater complexity, though BIC applies a stronger penalty, especially with larger datasets. To evaluate the relative performance of models, we compute the differences in AIC and BIC as \( \Delta X = \Delta \text{AIC} \) or \( \Delta X = \Delta \text{BIC} \). These differences are interpreted as follows:
\begin{itemize}
\item \textbf{$0 \leq \Delta X \leq 2$}:The evidence is \textit{weak} and it is impossible to judge whether model is superior.
\item \textbf{$2 < \Delta X \leq 6$}: Evidence is \textit{positive} in support of the model with the lower value. 
\item \textbf{$6 < \Delta X \leq 10$}: Evidence is considered to be \textit{strong}.
\end{itemize}
These statistical diagnostics serve as essential tools for model comparison, ensuring both accuracy and parsimony in cosmological inference.

\section{Constraints and Cosmological Implications}\label{sec5}
Using the observational data and methodology as mention in the above section, we investigate the constraints and cosmological implications for the two cases as mentioned in Sec. (\ref{sec3}). The results indicated by Table \ref{tab:full_stats_comparison} shows that both the models yields lower values of $\chi^2_{\min}$, AIC, and BIC compared to the standard $\Lambda$CDM model for the respective observational data under consideration.  Moreover, the differences in information criteria ($\Delta$AIC and $\Delta$BIC) consistently exceed 2 for all combinational dataset in case of CPL + $f(Q)$, providing positive evidence in favour of the model over $\Lambda$CDM. In case of BA + $f(Q)$ all information criteria consistently exceed 2, but for P-BAO+DESI dataset it's less then 2. The higher $R^2$ scores across all datasets further confirm an improved goodness of fit to observational data for both CPL + $f(Q)$ and BA + $f(Q)$ . These findings collectively demonstrate that both the model are capable of effectively capturing the DESI, P-BAO, and their combined datasets. And, under the light of the constraints as shown in Tables \ref{table:bestfits1} and \ref{table:bestfits3}, we investigate the respective cosmological dynamics as shown below.
\begin{figure*}[htb]
\centerline{
\includegraphics[width=.45\textwidth]{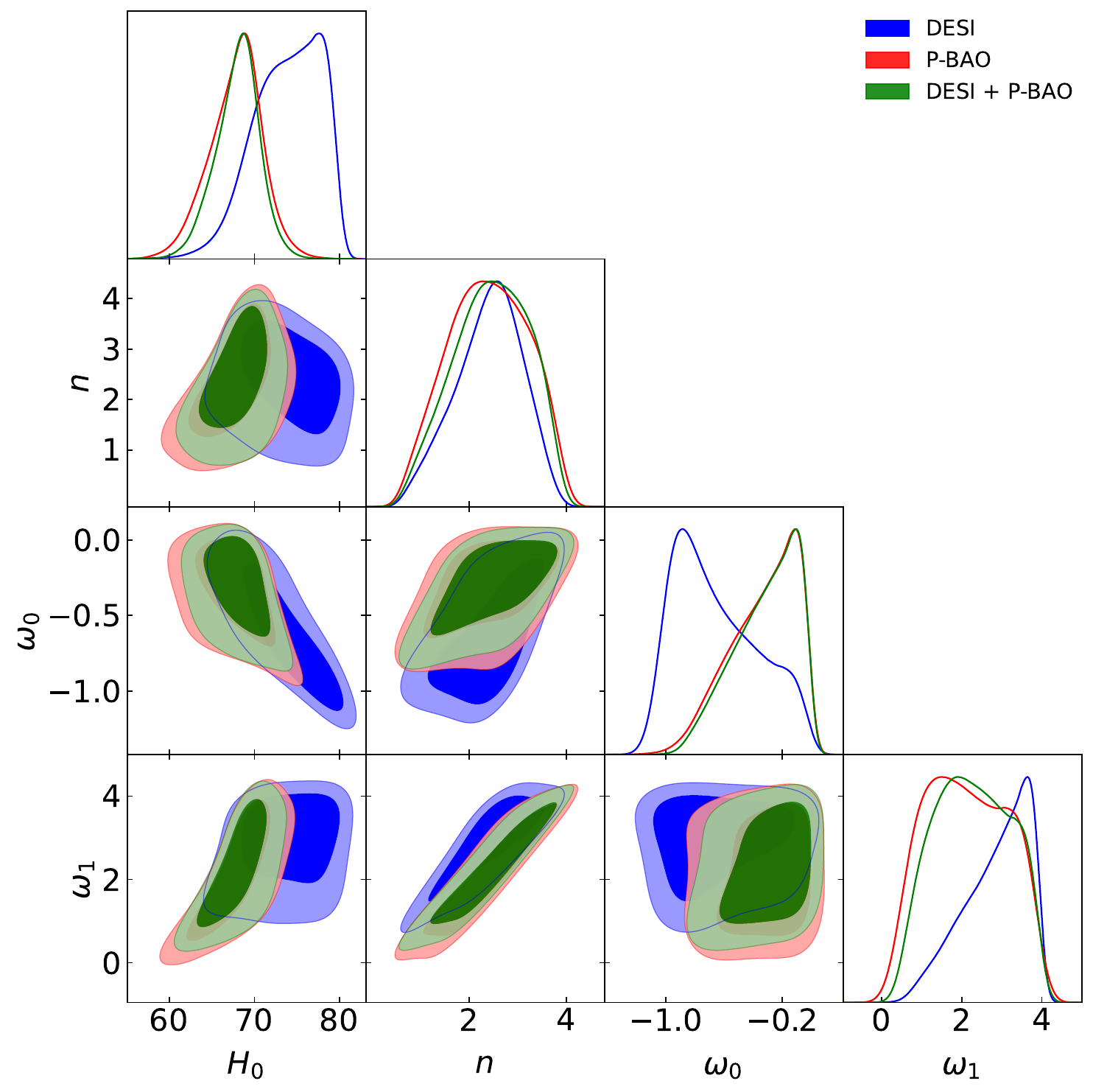}
\includegraphics[width=.45\textwidth]{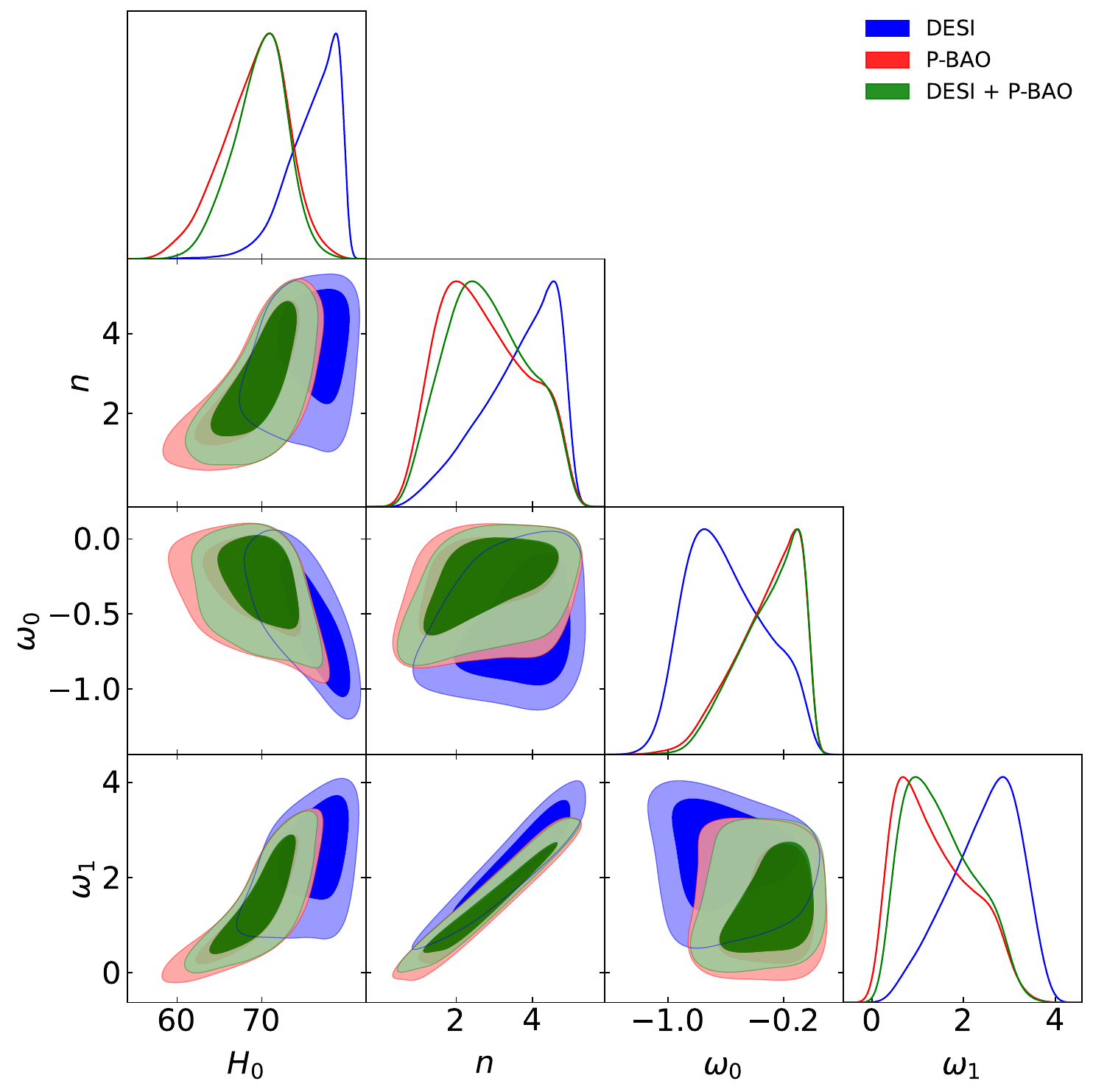}}
\caption{2-d contour sub-plot for the parameters $H_0$, $n$, $\omega_0$ and $\omega_1$ with 1-$\sigma$ and 2-$\sigma$ errors (showing the 68\% and 95\% c.l.) for $H(z)$ vs $z$. Right plot: CPL + $f(Q)$ and left plot: BA+ $f(Q)$.}
\label{figHC1}
\end{figure*}
\begin{figure*}[htb]
\centerline{
\includegraphics[width=.5\textwidth]{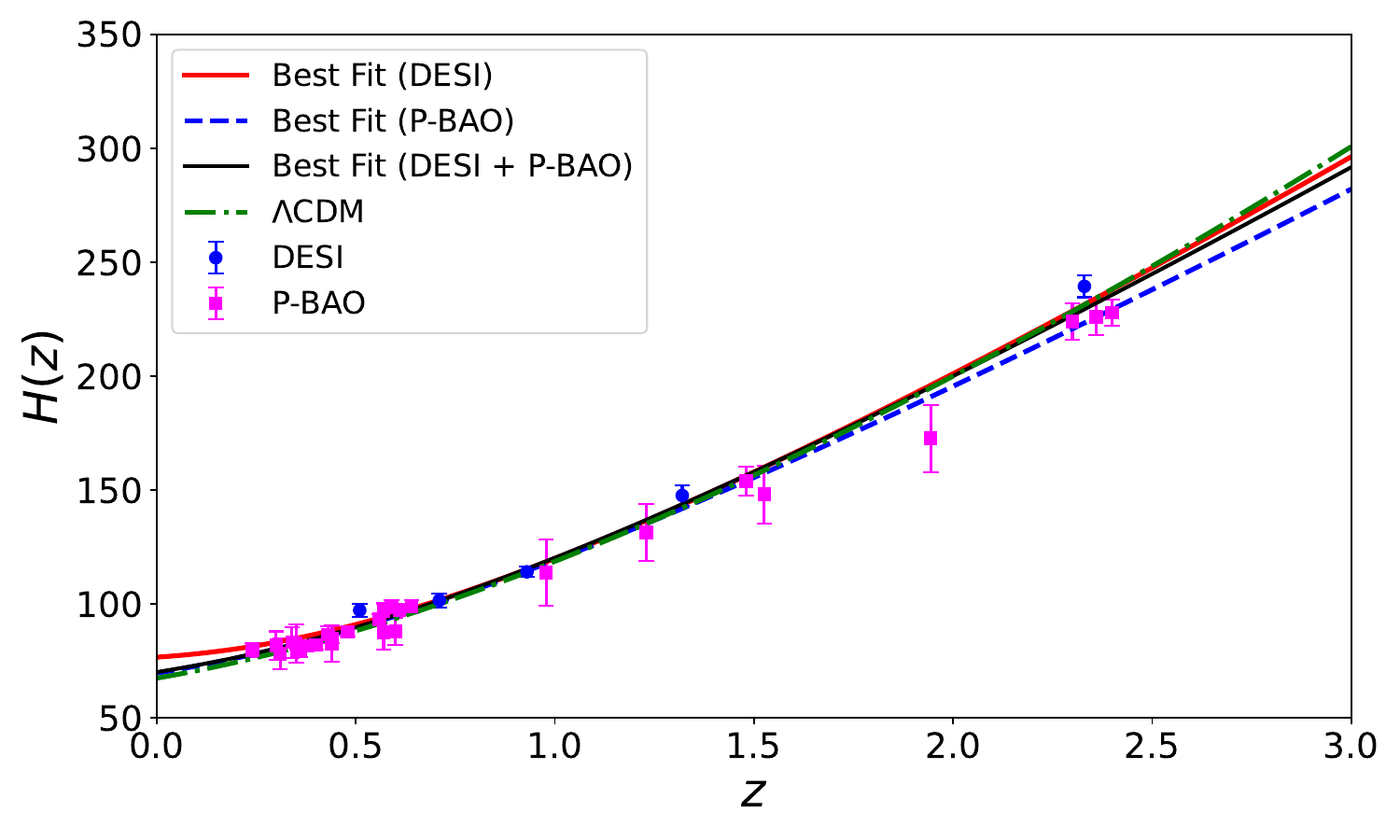}
\includegraphics[width=.5\textwidth]{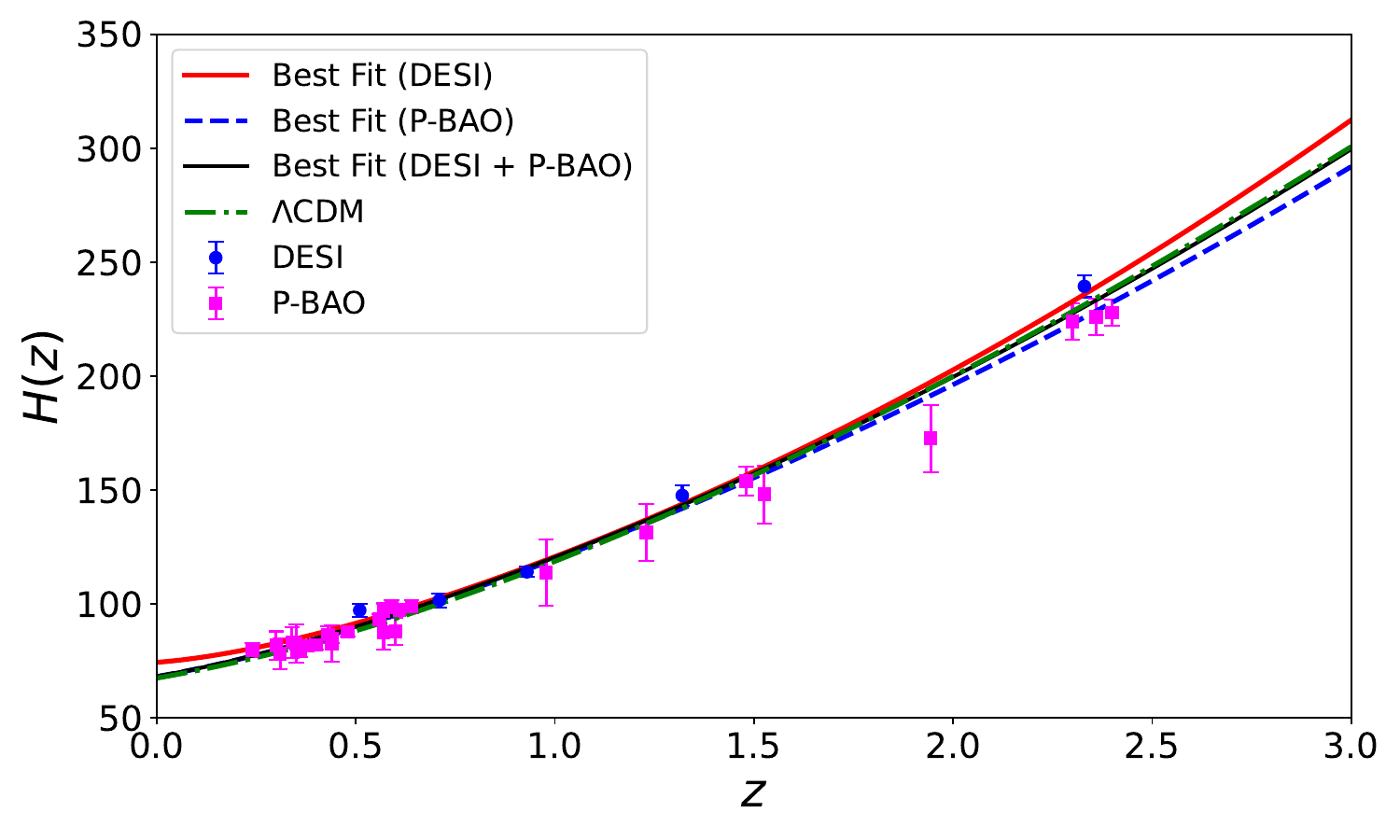}}
\caption{Plot of $H(z)$ vs $z$ for the best fit of the model parameters against the observational data. Here, a comparison is made with the $\Lambda$CDM model. Right plot: CPL + $f(Q)$ and left plot: BA+ $f(Q)$.}
\label{figHC2}
\end{figure*}

\begin{table*}[htb]
\centering
\renewcommand{\arraystretch}{1.2}
\begin{tabular}{llcccccc}
\toprule
\textbf{Model} & \textbf{Dataset} & \textbf{$R^2$} & \textbf{$\chi^2_{\min}$} & \textbf{AIC} & \textbf{BIC} & \textbf{$\Delta$AIC} & \textbf{$\Delta$BIC} \\
\midrule
\multirow{3}{*}{\textbf{CPL+$f(Q)$}} 
  & DESI           & 0.9957 & 5.62  & 13.62 & 12.06 & 5.96  & 6.35 \\
  & P-BAO          & 0.9885 & 14.17 & 22.17 & 27.35 & 4.86  & 3.56 \\
  & DESI+P-BAO     & 0.9849 & 29.09 & 37.09 & 42.96 & 3.51  & 2.04 \\
\midrule
\multirow{3}{*}{\textbf{BA+$f(Q)$}} 
  & DESI           & 0.9917 & 8.09  & 16.09 & 14.53 & 3.49  & 3.88 \\
  & P-BAO          & 0.9893 & 14.17 & 22.17 & 27.35 & 4.86  & 3.56 \\
  & DESI+P-BAO     & 0.9848 & 29.76 & 37.76 & 43.62 & 2.84  & 1.38 \\
\midrule
\multirow{3}{*}{\textbf{$\Lambda$CDM}} 
  & DESI           & 0.9878 & 13.58 & 19.58 & 18.41 & --    & --   \\
  & P-BAO          & 0.9830 & 21.03 & 27.03 & 30.91 & --    & --   \\
  & DESI+P-BAO     & 0.9846 & 34.60 & 40.60 & 45.00 & --    & --   \\
\bottomrule
\end{tabular}
\caption{Statistical comparison of the CPL+$f(Q)$ and BA+$f(Q)$ models with $\Lambda$CDM using DESI, P-BAO, and combined $H(z)$ datasets. $\Delta$AIC and $\Delta$BIC are calculated as $\Lambda$CDM $-$ Model.}
\label{tab:full_stats_comparison}
\end{table*}

\begin{table*}[ht]
\centering
\renewcommand{\arraystretch}{1.2}
\begin{tabular}{lcccc}
\toprule
\textbf{Dataset} & $\boldsymbol{H_0}$ & $\boldsymbol{n}$ & $\boldsymbol{\omega_0}$ & $\boldsymbol{\omega_1}$ \\
\midrule
\textbf{DESI} 
  & $74.33^{+3.86}_{-4.46}$ 
  & $2.45^{+0.64}_{-0.79}$ 
  & $-0.70^{+0.41}_{-0.25}$ 
  & $3.04^{+0.69}_{-1.06}$ \\
  
\textbf{P-BAO} 
  & $68.09^{+2.60}_{-3.43}$ 
  & $2.40^{+0.89}_{-0.88}$ 
  & $-0.29^{+0.20}_{-0.30}$ 
  & $2.10^{+1.24}_{-1.11}$ \\

\textbf{DESI + P-BAO} 
  & $68.20^{+2.21}_{-2.77}$ 
  & $2.48^{+0.81}_{-0.86}$ 
  & $-0.27^{+0.19}_{-0.28}$ 
  & $2.26^{+1.11}_{-1.02}$ \\
\bottomrule
\end{tabular}
\caption{Best-fit parameters for the \textbf{CPL+$f(Q)$} model with $1\sigma$ uncertainties for each $H(z)$ dataset.}
\label{table:bestfits1}
\end{table*}

\begin{table*}[ht]
\centering
\renewcommand{\arraystretch}{1.2}
\begin{tabular}{lcccc}
\toprule
\textbf{Dataset} & $\boldsymbol{H_0}$ & $\boldsymbol{n}$ & $\boldsymbol{\omega_0}$ & $\boldsymbol{\omega_1}$ \\
\midrule
\textbf{DESI} 
  & $76.62^{+2.43}_{-3.62}$ 
  & $3.83^{+0.84}_{-1.29}$ 
  & $-0.62^{+0.34}_{-0.25}$ 
  & $2.54^{+0.67}_{-0.94}$ \\

\textbf{P-BAO} 
  & $69.60^{+3.27}_{-4.41}$ 
  & $2.60^{+1.45}_{-1.06}$ 
  & $-0.27^{+0.19}_{-0.29}$ 
  & $1.23^{+1.10}_{-0.72}$ \\

\textbf{DESI + P-BAO} 
  & $70.05^{+2.77}_{-3.53}$ 
  & $2.77^{+1.29}_{-1.04}$ 
  & $-0.26^{+0.18}_{-0.27}$ 
  & $1.39^{+0.97}_{-0.70}$ \\
\bottomrule
\end{tabular}
\caption{Best-fit parameters for the \textbf{BA+$f(Q)$} model with $1\sigma$ uncertainties for each $H(z)$ dataset.}
\label{table:bestfits3}
\end{table*}

\subsection{Deceleration Parameter}
\begin{figure*}[htb]
\centerline{
\includegraphics[width=1\textwidth]{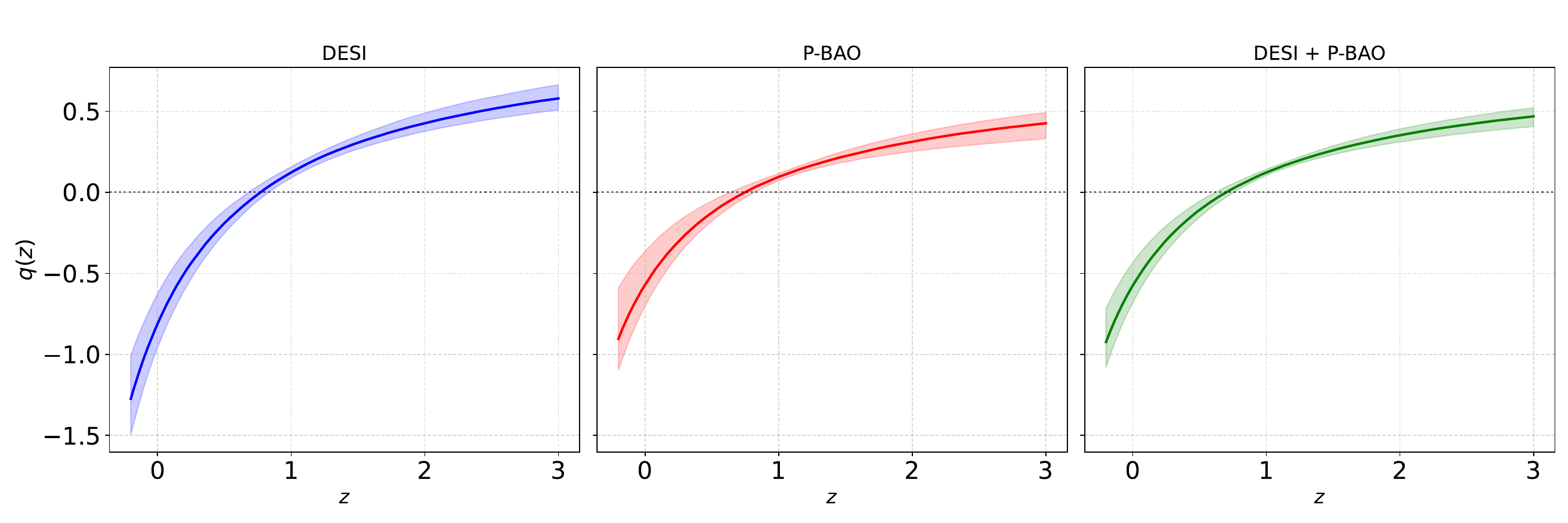}}
\centerline{
\includegraphics[width=1\textwidth]{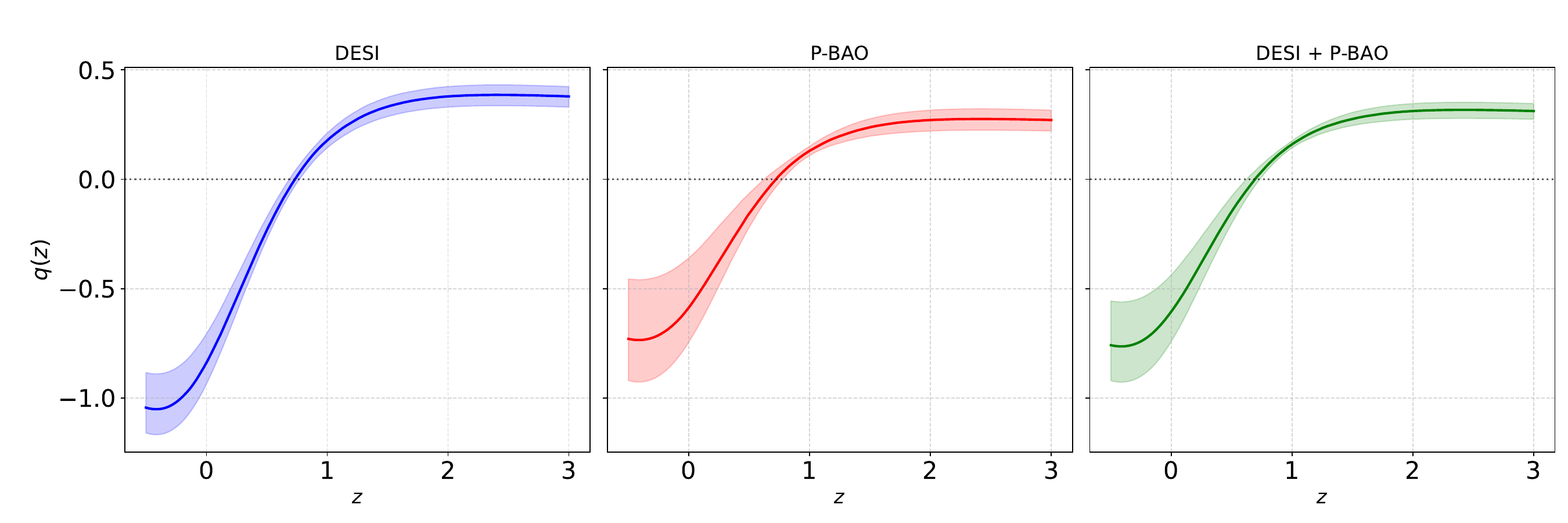}}
\caption{Evolution of deceleration parameter with redshift for for different datasets. The shaded regions here indicate the allowed region at 1$\sigma$ confidence level. Top plot: CPL + $f(Q)$ and bottom plot: BA+ $f(Q)$.}
\label{figHC3}
\end{figure*}
The deceleration parameter $q(z)$ is crucial for understanding the nature of cosmic expansion. A positive $q$ implies deceleration (as in a matter- or radiation-dominated era), whereas a negative $q$ indicates cosmic acceleration driven by dark energy. In our models, the evolution of $q(z)$ clearly demonstrates a transition from a decelerated to an accelerated phase.
From the MCMC analysis using combined datasets, we find that the present value of the deceleration parameter lies in the range $-1 < q(0) < 0$, consistent with an accelerating Universe. The exact transition redshift, $z_{\text{tr}}$, at which the Universe switches from deceleration to acceleration, falls within a moderate range (typically $z_{\text{tr}} \sim 0.7$), in line with standard observations. Details for which are illustrated by Fig.~\ref{figHC3} and Table~\ref{t5}. This dynamic evolution supports the idea that the $f(Q)$ model, with the adopted dark energy parameterizations, effectively captures the late-time acceleration without invoking a cosmological constant explicitly. The value of $q(0)$ suggests a stronger acceleration in some datasets. In particular, the DESI dataset tends to favor slightly stronger acceleration, with $q(0)$ values deeper in the negative regime. Both parameterizations predict similar redshift ranges for the onset of acceleration, though BA + $f(Q)$ generally exhibits more pronounced acceleration at $z = 0$.

\begin{table}[ht]
    \centering
    \caption{Best-fit values of the present-day deceleration parameter $q(0)$ and the transition redshift $z_{\rm tr}$ for CPL and BA parameterizations in $f(Q)$ gravity, using DESI, P-BAO, and combined datasets.}
    \label{t5}
    \begin{tabular}{|c|c|c|c|}
        \hline
        \textbf{Model} & \textbf{Dataset} & \boldmath$q(0)$ & \boldmath$z_{\rm tr}$ \\
        \hline
        {CPL + $f(Q)$} 
        & DESI & $-0.7990^{+0.1764}_{-0.1614}$ & $0.7639^{+0.0679}_{-0.0888}$ \\
        & P-BAO & $-0.5760^{+0.1985}_{-0.1321}$ & $0.7524^{+0.0613}_{-0.1121}$ \\
        & DESI + P-BAO & $-0.5746^{+0.1448}_{-0.1094}$ & $0.7017^{+0.0560}_{-0.0841}$ \\
        \hline
        {BA + $f(Q)$} 
        & DESI & $-0.8394^{+0.1337}_{-0.1033}$ & $0.7348^{+0.0321}_{-0.0525}$ \\
        & P-BAO & $-0.5889^{+0.2266}_{-0.1594}$ & $0.7203^{+0.0543}_{-0.0916}$ \\
        & DESI + P-BAO & $-0.6018^{+0.1780}_{-0.1277}$ & $0.6874^{+0.0465}_{-0.0678}$ \\
        \hline
    \end{tabular}
\end{table}

\subsection{Effective EoS Parameter}
\begin{figure*}[htb]
\centerline{
\includegraphics[width=1\textwidth]{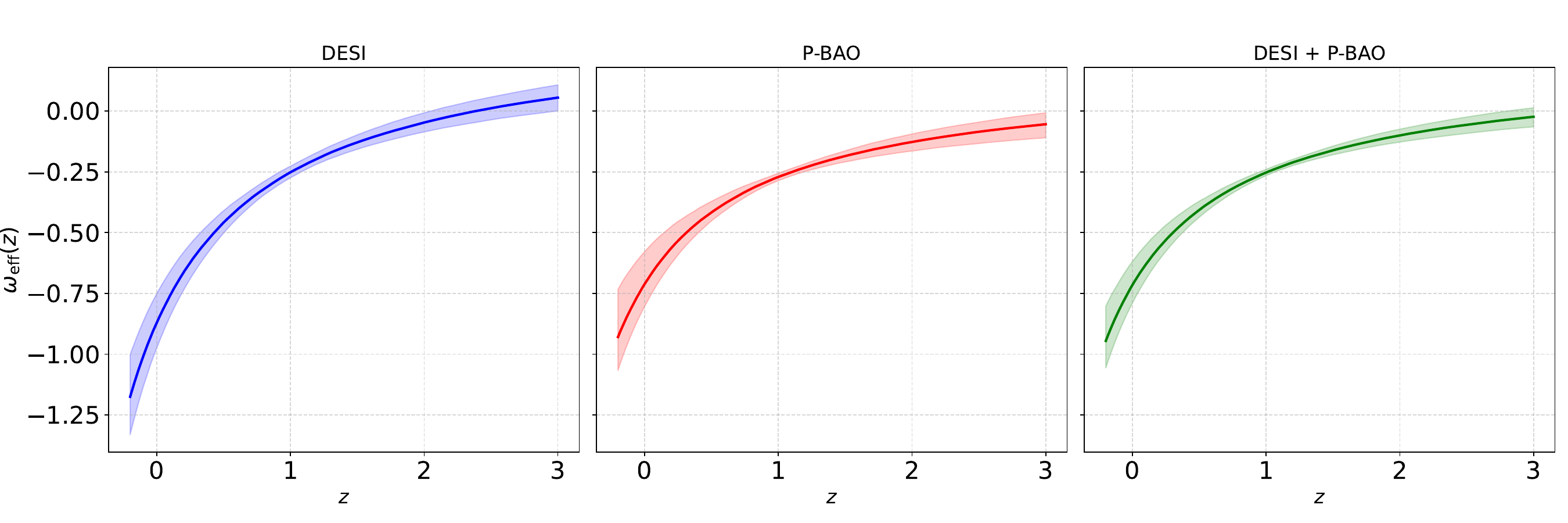}}
\centerline{
\includegraphics[width=1\textwidth]{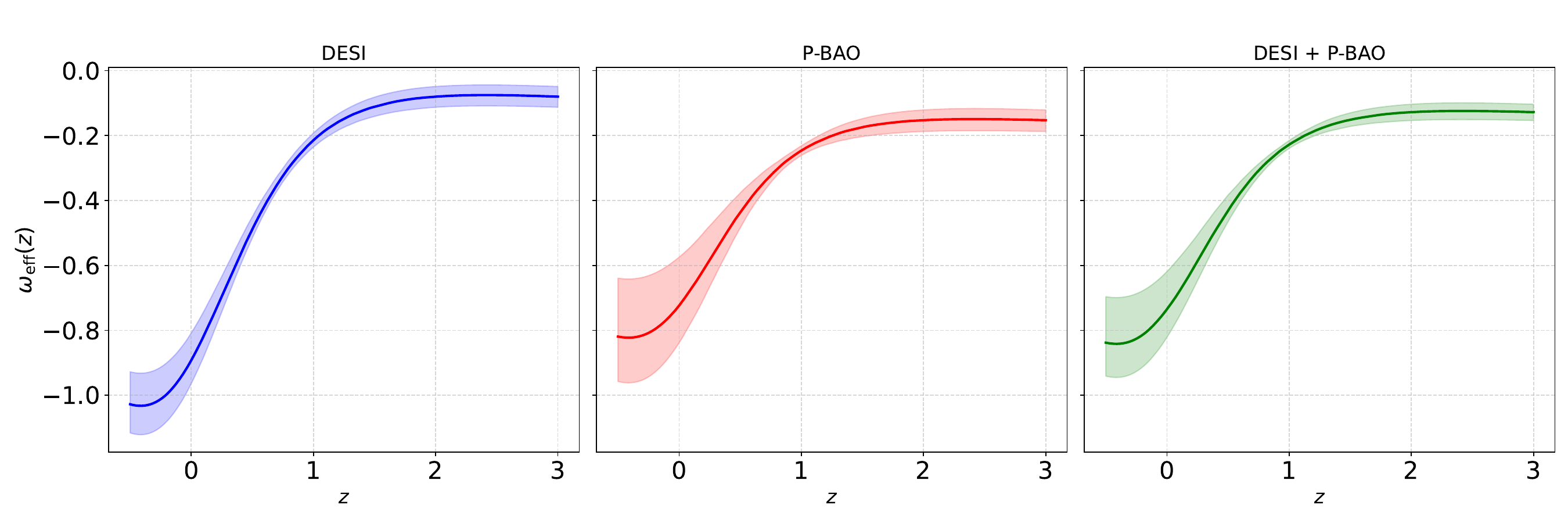}}
\caption{Evolution of effective EoS parameter with redshift for for different datasets. The shaded regions here indicate the allowed region at 1$\sigma$ confidence level. Top plot: CPL + $f(Q)$ and bottom plot: BA+ $f(Q)$.}
\label{figHC4}
\end{figure*}
The effective equation of state (EoS) parameter, $\omega_{\rm eff}(z)$, reflects the total dynamical behavior of the cosmic fluid. It captures the combined effects of matter, dark energy, and any modified gravity contributions. Values of $\omega_{\rm eff}(z) < -1/3$ signify an accelerating Universe, while $\omega_{\rm eff}(z) = -1$ corresponds to a cosmological constant ($\Lambda$CDM). More negative values ($\omega_{\rm eff}(z) < -1$) indicate a phantom-like dark energy component. Additionally, there are even other accelerating phases of the Universe also, which are characterized by an effective equation of state parameter $\omega < -1/3$. This phase includes both the quintessence regime ($-1 < \omega \leq -1/3$) and the phantom regime ($\omega < -1$). Our analysis reveals that all models yield $-1 < \omega_{\rm eff}(0) <  -1/3$, confirming the presence of late-time acceleration phase which is quintessence-like across all datasets for the present time.
Notably, the DESI dataset leads to deeper negative values of $\omega_{\rm eff}(0)$ in both parametrizations, with the BA + $f(Q)$ model suggesting a stronger acceleration compared to CPL + $f(Q)$. The effective EoS behavior over redshift is visualized in Fig.~\ref{figHC4}, while Table~\ref{tab:weff} summarizes the best-fit values of $\omega_{\rm eff}(0)$ at $z = 0$. These results reinforce the conclusion that the current Universe is in a phase of accelerated expansion. For both CPL and BA forms, the DESI dataset pushes the EoS deeper into the phantom regime for lower $z$ values, highlighting the influence of the underlying parameterization and data precision. However, other datasets for BA particularly appears more consistent with quintessence-like behaviour, and for CPL we find it mildly phantom-like for lower redshift values. For more insights we hence perform the $Om$ diagnostics.
\begin{table}[ht]
    \centering
    \caption{Best-fit values of the present-day effective EoS parameter $\omega_{\rm eff}(0)$ for CPL and BA parameterizations in $f(Q)$ gravity, using DESI, P-BAO, and combined datasets.}
    \label{tab:weff}
    \begin{tabular}{|c|c|c|}
        \hline
        \textbf{Model} & \textbf{Dataset} & \boldmath$\omega_{\rm eff}(0)$ \\
        \hline
        {CPL + $f(Q)$} 
        & DESI & $-0.8660^{+0.1176}_{-0.1076}$ \\
        & P-BAO & $-0.7173^{+0.1323}_{-0.0881}$ \\
        & DESI + P-BAO & $-0.7164^{+0.0965}_{-0.0729}$ \\
        \hline
        {BA + $f(Q)$} 
        & DESI & $-0.8929^{+0.0891}_{-0.0688}$ \\
        & P-BAO & $-0.7259^{+0.1511}_{-0.1063}$ \\
        & DESI + P-BAO & $-0.7346^{+0.1187}_{-0.0851}$ \\
        \hline
    \end{tabular}
\end{table}
\subsection{$Om$ Diagnostics}
\begin{figure*}[htb]
\centerline{
\includegraphics[width=1\textwidth]{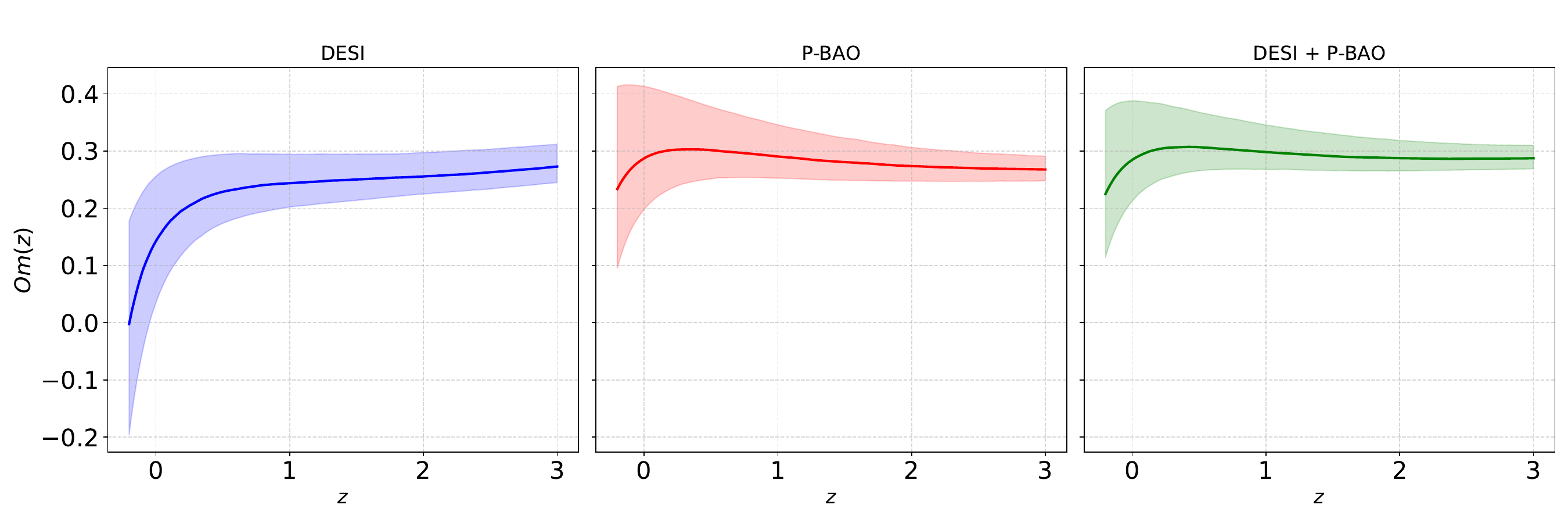}}
\centerline{
\includegraphics[width=1\textwidth]{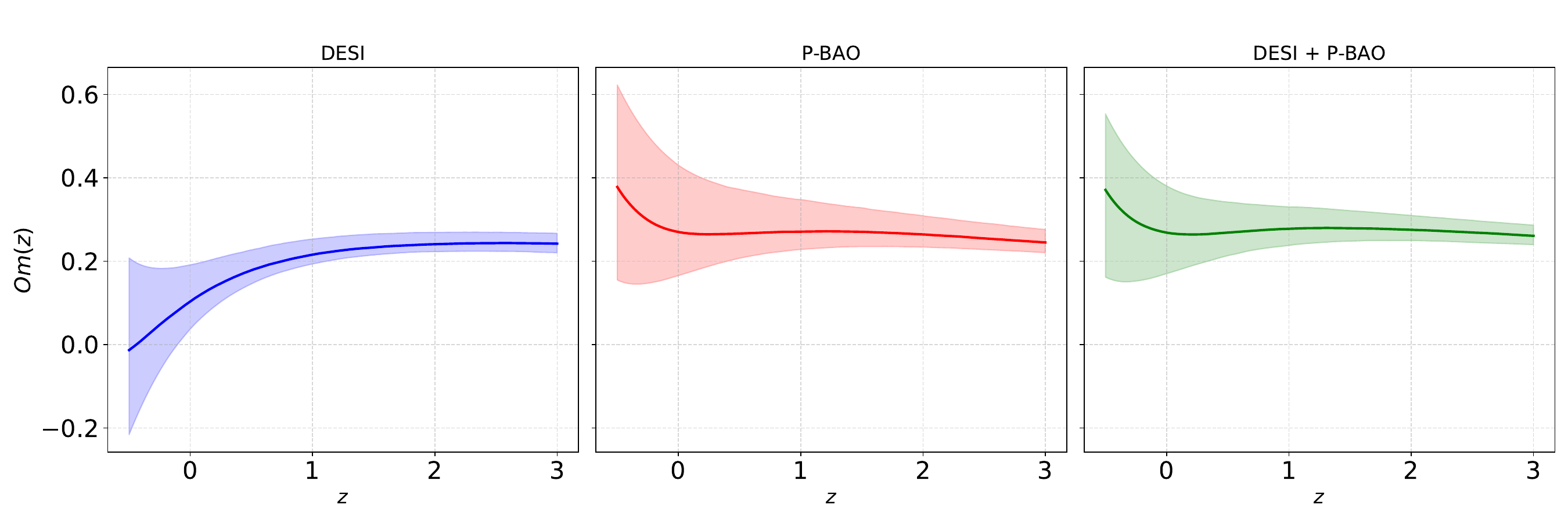}}
\caption{Evolution of $Om(z)$ diagnostics with redshift for for different datasets. The shaded regions here indicate the allowed region at 1$\sigma$ confidence level. Top plot: CPL + $f(Q)$ and bottom plot: BA+ $f(Q)$.}
\label{figHC5}
\end{figure*}
The $Om(z)$ diagnostic is a purely geometrical tool constructed from the Hubble parameter, allowing for discrimination among dark energy models without requiring knowledge of the equation of state. It is defined as:
\begin{equation}
Om(z) = \frac{H^2(z)/H_0^2 - 1}{(1 + z)^3 - 1}.
\end{equation}
In the standard $\Lambda$CDM model, $Om(z)$ remains constant with redshift and equals the matter density parameter $\Omega_{m0}$. Deviations from constancy signal departures from $\Lambda$CDM, with the direction of the slope offering physical insights:
\begin{itemize}
    \item \textbf{Flat $Om(z)$} indicates $\Lambda$CDM-like behavior.
    \item \textbf{Negative slope of $Om(z)$} implies quintessence-like dynamics ($\omega > -1$).
    \item \textbf{Positive slope of $Om(z)$} implies phantom-like behavior ($\omega < -1$).
\end{itemize}
Figure~\ref{figHC5} presents the $Om(z)$ curves constructed from MCMC samples for both CPL and BA parameterizations in $f(Q)$ gravity. The diagnostic reveals distinct features for each case:
\begin{itemize}
    \item For \textbf{CPL + $f(Q)$}, the best parameters according to all the datasets (.i.e. DESI, P-BAO and P-BAO+DESI ) shows $Om(z)$ to have a positive slope for smaller redshift values indicating at a phantom like behaviour.  
    \item For \textbf{BA + $f(Q)$}, the $Om(z)$ curves remain nearly flat across all datasets for higher redshift values, with only deviations at lower redshift values. The best parameters according to DESI dataset shows $Om(z)$ to have a positive slope for smaller redshift values indicating at a phantom like behaviour, however other datasets like P-BAO and P-BAO+DESI shows the negative slope for smaller redshift values indicating at a quintessence-like behaviour.
\end{itemize}
These results align with the analysis of the deceleration and effective EoS parameters and further highlight the sensitivity of diagnostics to the dark energy parameterization.

\section{Conclusion}\label{sec6}
In this work, we have explored the cosmological implications of symmetric teleparallel gravity through the power-law $f(Q) = \alpha Q^n$ model, combined with two phenomenologically distinct dark energy parametrizations: Chevallier–Polarski–Linder (CPL) and Barboza–Alcaniz (BA). Using updated BAO measurements from both pre-DESI and the recently released DESI DR2 datasets, we constrained the model parameters via MCMC and analyzed key cosmological quantities such as $q(z)$, $\omega_{\rm eff}(z)$, and Om diagnostics.\\
Statistical comparisons with the $\Lambda$CDM model using $\chi^2$, AIC, BIC, and $R^2$ metrics show that both $f(Q)$ parametrizations offer competitive or superior fits. The $\Delta$AIC and $\Delta$BIC values for CPL + $f(Q)$ exceed the critical threshold of 2 across most datasets, indicating positive evidence against $\Lambda$CDM. These results reinforce the ability of $f(Q)$ gravity to mimic or outperform the concordance model in terms of fit and dynamical flexibility. Further analysis reveals a consistent acceleration phase with $q(0) < 0$ and $-1<\omega_{\rm eff}(0) < -1/3$, supporting quintessence-like expansion for the present time. In both CPL and BA parametrizations within the $f(Q)$ framework, the DESI dataset drives the effective equation of state (EoS) deeper into the phantom regime at low redshifts, emphasizing the impact of both the EoS form and the precision of observational data. For the BA case, while DESI favours a more phantom-like behaviour, other datasets such as P-BAO and the combined P-BAO+DESI suggest a quintessence-like regime at present. The behaviour of the $Om(z)$ diagnostic further reinforces this trend. In BA + $f(Q)$, $Om(z)$ remains nearly constant at higher redshifts across all datasets—mimicking $\Lambda$CDM—but begins to diverge at lower redshifts. Specifically, DESI favors a slight positive slope in $Om(z)$ at low $z$, hinting at phantom-like evolution, whereas P-BAO and the combined datasets exhibit a negative slope, consistent with quintessence-like dynamics. In contrast, the CPL + $f(Q)$ model demonstrates greater consistency: all three datasets (DESI, P-BAO, and DESI + P-BAO) yield a positively sloped $Om(z)$ at low redshift, pointing toward a mild but persistent phantom-like character in the late-time expansion.\\
A central outcome of this study is the demonstrable influence of the new DESI DR2 BAO dataset on the inferred cosmic dynamics within the $f(Q)$ framework. While earlier BAO data already supported an accelerated expansion, the inclusion of DESI data leads to tighter constraints and shifts the interpretation of key indicators, particularly at low redshifts. Notably, DESI pushes the effective equation of state parameter $\omega_{\rm eff}(z)$ deeper into the phantom regime, suggesting stronger late-time acceleration than previously inferred. Similarly, the deceleration parameter $q(z)$ shows a steeper decline at recent times, and the transition redshift $z_{\text{tr}}$ moves slightly earlier, reinforcing the picture of an earlier onset of acceleration. The $Om(z)$ diagnostic—traditionally flat for $\Lambda$CDM—shows a distinct positive slope at low $z$ in the DESI-only case, pointing toward phantom-like tendencies, whereas prior datasets suggested quintessence-like or near-$\Lambda$CDM behavior depending on the model. These shifts highlight how higher-precision, larger-volume observations such as DESI DR2 can meaningfully reshape our understanding of late-time cosmic expansion, even within the same theoretical framework. The results emphasize that conclusions about the nature of dark energy and the evolution of the Universe are increasingly data-driven and sensitive to the quality and scale of observations. As data quality and volume continue to improve, especially with ongoing and future large-scale structure surveys like DESI, the synergy between advanced parametrization techniques and alternative gravity theories will be instrumental in discriminating between competing scenarios. This study reaffirms that the evolving landscape of observational cosmology demands continual reassessment of theoretical frameworks, leveraging both novel parametrizations and modifications of gravity to fully decode the Universe's accelerating expansion. However, in this work a limited portion of the DESI DR2 dataset was used, in future we aim to add more larger and diverse dataset to further draw conclusions on novel parametrization schemes and modified gravity theories.

\end{document}